\documentclass[superscriptaddress,prd,aps,preprint]{revtex4}

\def\tr{\mathrm{tr}}
\def\sech{\mathrm{sech}}
\def\1{{\bf 1}}
\newcommand{\be}{\begin{equation}}
\newcommand{\en}{\end{equation}}
\newcommand{\ba}{\begin{eqnarray}}
\newcommand{\ea}{\end{eqnarray}}

\newcommand{\Slash}[1]{\!\not\!{#1}}
\newcommand{\SLASH}[1]{\not\!\!{#1}}

\begin{document}

\title{Anomaly cancellation in three-dimensional noncommutative gauge theories}

\author{M. Gomes}
\author{T. Mariz}
\affiliation{Instituto de F\'{\i}sica, Universidade de S\~ao Paulo\\
Caixa Postal 66318, 05315-970, S\~ao Paulo, SP, Brazil}
\email{mgomes,tmariz,ajsilva@fma.if.usp.br}
\author{J. R. Nascimento} 
\affiliation{Instituto de F\'{\i}sica, Universidade de S\~ao Paulo\\
Caixa Postal 66318, 05315-970, S\~ao Paulo, SP, Brazil}
\affiliation{Departamento de F\'{\i}sica, Universidade Federal da Para\'{\i}ba\\
 Caixa Postal 5008, 58051-970, Jo\~ao Pessoa, Para\'{\i}ba, Brazil}
\email{jroberto,petrov,edilberto@fisica.ufpb.br}
\author{A. Yu. Petrov}
\affiliation{Departamento de F\'{\i}sica, Universidade Federal da Para\'{\i}ba\\
 Caixa Postal 5008, 58051-970, Jo\~ao Pessoa, Para\'{\i}ba, Brazil}
\email{jroberto,passos,petrov@fisica.ufpb.br}
\author{A. J. da Silva}
\affiliation{Instituto de F\'{\i}sica, Universidade de S\~ao Paulo\\
Caixa Postal 66318, 05315-970, S\~ao Paulo, SP, Brazil}
\email{mgomes,ajsilva@fma.if.usp.br}
\author{E. O. Silva}
\affiliation{Departamento de F\'{\i}sica, Universidade Federal da Para\'{\i}ba\\
 Caixa Postal 5008, 58051-970, Jo\~ao Pessoa, Para\'{\i}ba, Brazil}
\email{jroberto,petrov,edilberto@fisica.ufpb.br}

\begin{abstract}
The anomaly found by Callan and Harvey is shown to be cancelled in a three-dimensional noncommutative gauge theory coupled to a fermion with a mass function depending on one spatial coordinate (domain wall mass). This evaluation has been done for the fermion in the fundamental and adjoint representations of the gauge group in the limit of small noncommutativity $\theta$ parameter.
\end{abstract}

\maketitle 

\section{Introduction}
Anomalies arise when some of the symmetries of { a} classical { model} fail in the quantum mechanics vacuum generating functional. If such a symmetry is a gauge symmetry associated with a dynamical gauge field { the model 
fails to satisfy the gauge principle}. The usual method  to { improve the model and restore gauge symmetry is by} 
the { introduction} of extra fermion species with appropriate quantum numbers. Following these ideas Callan and Harvey 
showed in 1984 \cite{CH} that the mathematical relation between gauge anomalies in $2n$ dimensions, the parity anomalies in $2n +1$ dimensions and the Dirac 
index density in $2n+2$ dimensions can be understood in terms of  the physics of fermionic zero modes on strings and domain walls  (see \cite{Naculich} for a detailed study of string aspects of this correspondence).
One of the interesting examples of application of this relation is given in \cite{Chand} where a $2+1$ dimensional theory of fermions with a domain wall mass, coupled to a gauge field is described. Zero mode chiral fermions appear in the $1+1$ dimensional domain wall and a breakdown of the gauge symmetry is induced. On the other hand, a three-dimensional gauge anomaly is generated by the nontrivial dependence of the mass on the (second) space coordinate, that exactly cancel the one induced by the chiral modes attached to the domain wall.
The  characteristics of these chiral modes have been explored in 
several  papers in the literature \cite{FosTor,Fos,dw}.
{ Further, the Callan-Harvey mechanism was also found to have important applications in the context of the condensed matter physics \cite{fen}.

The high interest on the noncommutative field theories (NCFTs), which have been intensively studied during the last years (for some reviews see \cite{NCFT}), opens the necessity of the study of the anomalies in this new context. One of 
the main characteristics of noncommutative field theories is the UV/IR mixing \cite{Minw} implying in highly nontrivial low-energy dynamics. A natural question in this setting is whether this nontrivial low-energy dynamics could generate new anomalies, which have no analogs in the commutative field theories, or modify the results about the previously known ones. Some important results in this direction are the calculation of the chiral anomaly in the two-dimensional noncommutative Schwinger model \cite{Pres}, the study of the axial \cite{Ard} and the chiral \cite{Mart} anomalies in the four-dimensional noncommutative QED and the study of the consistent chiral anomalies in the four-dimensional Yang-Mills theories \cite{Bon}. In spite of these investigations it still remains to understand the Callan-Harvey effect. This is precisely the
 aim of this paper, i.e., to extend the study 
 carried on in \cite{Chand}, to the noncommutative 
situation.
The main object of our study is a three-dimensional model of a fermion
field, with a domain wall mass, coupled to a gauge field in a noncommutative way. We show that the
gauge anomalies induced by the zero mode chiral fermions in even (in this case two) dimensions, are 
cancelled by
the ones induced by the parity-odd mass term in the next higher (three) dimensions.

The structure of the paper is the following. In the Section \ref{section2} we describe the action and the propagators of the theory. The section \ref{section3} is devoted to the study of the two-point vertex function of the gauge field, both in the fundamental and adjoint representations of the gauge group and its possible contributions to the anomaly which is shown to be exactly cancelled. In the Summary the results are discussed.

\section{Action of the theory}\label{section2}

We start with the following three-dimensional action of the spinor field coupled to the external gauge field in the fundamental and adjoint representations, which is a known spinor sector of the noncommutative QED \cite{NCQED}:
\be
S=\int d^3z\,\bar{\psi}\star\left[i\gamma^\mu D_\mu\psi + m(s)\star\psi\right],
\en
where the covariant derivative is given by
\begin{eqnarray}\label{not}
D_\mu\psi = \left\{ \begin{array}{ll}
                      \partial_{\mu}\psi - ieA_{\mu}\star\psi,
& \;\mbox{for the fundamental representation},\\
                      \partial_{\mu}\psi - ie[A_{\mu},\psi]_\star,
& \;\mbox{for the adjoint representation,}
\end{array}
\right.
\end{eqnarray}
and the star indicates the Moyal product
\begin{equation} \label{produtoMoyal}
\phi_{1}(x)\star\phi_{2}(x)=\lim_{y\rightarrow x}e^{\frac{i}{2}\theta^{\mu
\nu}\frac{\partial}{\partial y^\mu}\frac{\partial}{\partial x^\mu}}\phi_{1}(y)\phi_{2}(x).
\end{equation}

This model is a natural noncommutative generalization of the one studied in \cite{CH,Chand}.
Here the coordinates are $z^{\mu}=(t,x,s)$, the signature is $\mathrm{diag}(+--)$, and the mass depends only on the second spacial dimension $s$. In general, in the fundamental representation the one-loop contribution to the two-point vertex function of the gauge field does not depend on the noncommutativity parameter and coincides with the commutative result because of the structure of the interaction vertex. On the other hand, in the adjoint representation case the nontrivial dependence on the noncommutativity parameter is always present.

In this study, for sake of simplicity we assume that $\theta$ is small in order to avoid affecting the mass by the noncommutativity. Smallness of $\theta$ is based on the fact that the noncommutativity can be naturally treated as a perturbation for the Schroedinger equation with the domain wall mass term, $m(s)\star\psi = m\left(s-\textstyle{i\over2}\theta^{2\mu} \partial_\mu\right)\psi$  $\sim m(s)\psi$ as has been studied in \cite{Chand}. Otherwise the only way of reasonable studying the theory is to rearrange the action, with the $m(s)$ term belonging to the interaction part, being treated as a coupling of the spinor field with some extra scalar field $m(s)$. Therefore, we propose $m(s)$ to be a function with the domain wall shape:
\be
m(s)=m_0\tanh(m_0 s),
\en
which provides $m(s)\to \pm\; m_0$ for $s\to \pm \;\infty$ and $m(s) \to 0$ for $s=0$. The propagator of the spinor field is the same in commutative and noncommutative cases and is given by (see \cite{Chand} for details)
\be\label{propf}
S(z,z') = \int\frac{d^3k}{(2\pi)^3} S(k;s,s')\, e^{-ik\cdot(z-z')},
\en
where $S(k;s,s')=S_1(k;s,s')+S_2(k;s,s')$, with
\be
S_1(k;s,s') = \frac{\Slash{k}+M(k;s,s')}{k^2-m^2_0},\label{S1}
\en
being the massive part, whereas
\be\label{S2}
S_2(k;s,s') = \frac{m_0}{4}(1+i\gamma^2)\sech(m_0s)\sech(m_0s')\frac{\Slash{k}}{k^2}\delta(k_2)
\en
is the chiral one generated by the chiral mode. The $M(k;s,s')$ is a mass matrix of the form
\ba
M(k;s,s') = \left(\begin{array}{cc}
-m(s) & \;\;\;\;\; \frac{k_0+k_1}{k^2_2+m^2_0}\{m(s)m(s')+ik_2[m(s')-m(s)]-m^2_0\}\\
0 & -m(s')
\end{array}
\right).
\ea
In the next section we will use this propagator for the study of the anomaly cancellation.

\section{Two-point function of the gauge field}\label{section3}

The lowest order perturbative correction to the two-point function of the gauge field, in the fundamental representation, determining the potential anomaly is
\be
S_\mathrm{eff}[A]=\frac{e^2}{2}\int d^3zd^3z'\mathrm{tr}\left[A_{\mu}(z)*\gamma^{\mu}S(z,z')*A_{\nu}(z')\gamma^{\nu}*S(z',z)\right],
\en
or in terms of noncommutative phase factors,
\be\label{Sef2}
S_\mathrm{eff}[A]=\frac{e^2}{2}\int d^3zd^3z'\,e^{i\sum_{i<j}^3\left(\partial_{x_i}\wedge\partial_{x_j}+\partial_{y_i}\wedge\partial_{y_j}\right)}\,
\mathrm{tr} \SLASH{A}(x)S(x,y)\SLASH{A}(y)S(y,x)\Big|_{\{x_i=z,y_i=z'\}}, 
\en
where $\partial_{x_1}\wedge\partial_{x_2} = \frac i2\theta^{\mu\nu}\partial_{x_{1\mu}}\partial_{x_{2\nu}}$ and $\tr$ implies taking the trace over Dirac matrices. In the paper \cite{Chand} the commutative analog of this expression was obtained by using of the coordinate representation of the propagator. However, the Moyal product in the coordinate representation is an infinite series of space-time derivatives, and therefore the use of the coordinate representation turns out to be highly problematic. To avoid the need of expanding the Moyal product order by order in $\theta$ we will use Fourier expansion of the propagator. This calculation meets two specific difficulties. The first  is that different terms in its the propagator are defined in different dimensions of the space-time, and the second is the nontrivial dependence of the propagator on $s$. Thus, by using the Fourier expansion of the propagator (\ref{propf}) and the straightforward representations 
\be
A_{\mu}(z)=\int\frac{d^3p}{(2\pi)^3}{\tilde A_{\mu}(p)}\,e^{ip\cdot z},
\en
we get( from now on we omit the tilde of $A$)
\ba\label{Sef3}
S_\mathrm{eff}[A] &=& \frac{e^2}{2} \int \frac{d^3p}{(2\pi)^3}\frac{d^3p'}{(2\pi)^3} \int \frac{d^3k}{(2\pi)^3}\frac{d^3l}{(2\pi)^3} \,e^{ip\wedge k+i(k-p)\wedge\,l}e^{ip'\wedge k+i(k+p')\wedge\,l} \nonumber\\
&\times& \int d^3z e^{-iz\cdot(k-p-l)} \int d^3z' e^{-iz'\cdot(l-k-p')} \,\mathrm{tr} \SLASH{A}(p)S(k;s,s')\SLASH{A}(p')S(l;s,s').
\ea
As the propagators are $s$ and $s'$ dependent we can only build two-dimensional Dirac delta functions rather than three-dimensional ones 
\be\label{deltazz'}
\int \frac{d^3z}{(2\pi)^3}\,e^{-iz'\cdot(k-p-l)} \int \frac{d^3z'}{(2\pi)^3}\,e^{-iz'\cdot(l-k-p')} = \delta^3(k-p-l)\delta^3(l-k-p').
\en
Thus, the resulting expression has a very complicated structure, but we remind the reader that to obtain the anomaly we must consider these contributions in the $m_0\to\infty$ limit. 

First of all, we will consider the term involving the $S_1(k;s,s')$ and $S_1(l;s,s')$ propagators. In the $m_0\to\infty$ limit, it is sufficient to approximate $M(k;s,s')\sim m_0\1$ and $M(l;s,s')\sim m_0\1$, so that now the three-dimensional Dirac delta functions (\ref{deltazz'}) can be used, and the contribution to the expression (\ref{Sef3}) can be written as
\be
S_{11} = \frac{e^2}{2} \int\frac{d^3p}{(2\pi)^3} A_\mu(p)A_\nu(-p)\Pi_{11}^{\mu\nu},
\en
where in the fundamental representation the phase factors mutually cancel, and the polarization tensor
\be
\Pi^{\mu\nu}_{11} = \tr \int\frac{d^3k}{(2\pi)^3} \gamma^\mu\frac{\Slash{k}+m_0}{k^2-m^2_0}\gamma^\nu\frac{\Slash{k}-\Slash{p}+m_0}{(k-p)^2-m^2_0}
\en
exactly reproduces that one for the commutative analog of the theory \cite{Chand}. 

Now, we will concentrate ourselves on the adjoint representation where the impact of the noncommutativity is nontrivial. For us to take into account the adjoint representation in the effective action (\ref{Sef2}) we should make the exchange
\be
e^{i\sum\left(\partial_{x_i}\wedge\partial_{x_j}+\partial_{y_i}\wedge\partial_{y_j}\right)} \rightarrow 2i \sin\left(\sum\partial_{x_i}\!\wedge\partial_{x_j}\right)2i\sin\left(\sum\partial_{y_i}\!\wedge\partial_{y_j}\right),
\en
so that now the Eq. (\ref{Sef3}) becomes
\ba\label{Sef4}
S_\mathrm{eff}[A] &=& -2e^2 \int \frac{d^3p}{(2\pi)^3}\frac{d^3p'}{(2\pi)^3} \int \frac{d^3k}{(2\pi)^3}\frac{d^3l}{(2\pi)^3} \sin[p\!\wedge\! k+(k-p)\!\wedge\!\,l]\sin[p'\!\wedge\! k+(k+p')\!\wedge\!\,l] \nonumber\\
&\times& \int d^3z e^{-iz\cdot(k-p-l)}\int d^3z' e^{-iz'\cdot(l-k-p')}\,\mathrm{tr} \SLASH{A}(p)S(k;s,s')\SLASH{A}(p')S(l;s,s').
\ea
Therefore, the contribution to the effective action (\ref{Sef4}) that involves the $S_1(k;s,s')$ and $S_1(l;s,s')$ propagators, in the $m_0\to\infty$ limit, presents the following expression for the polarization tensor
\be\label{S11}
\Pi^{\mu\nu}_{11} = 4 \,\tr \int\frac{d^3k}{(2\pi)^3} \gamma^\mu\frac{\Slash{k}+m_0}{k^2-m^2_0}\gamma^\nu\frac{\Slash{k}-\Slash{p}+m_0}{(k-p)^2-m^2_0}\sin^2(p\!\wedge\! k). 
\en
As it is usual in the noncommutative field theory, we can split the above expression into the  planar and nonplanar parts by replacing $\sin^2(p\wedge k)=\frac{1}{2}[1-\cos(2p\wedge k)]$. Hence, the planar ($\theta$ independent) contributions differ from the commutative contributions by just the factor $\frac{1}{2}$. Thus, the nontrivial contribution to the anomaly can arise only from the nonplanar sector. The potential anomaly occurs in the terms of Eq. (\ref{S11}) which has a three-dimensional antisymmetric tensor $\epsilon^{\mu\nu\lambda}$. This is exactly the noncommutative Chern-Simons term that was calculated in \cite{Mariz}, given by
\be\label{CS}
S_\mathrm{CS} = -\frac {ie^2}{8\pi}\int d^3z \;\mathrm{sgn}(m_0)\left(1-e^{-m_0|\bar\theta|}\right)\epsilon^{\mu\nu\lambda}A_\mu \partial_\nu A_\lambda,
\en
in which we go back to the space of coordinates, where $\bar\theta^\mu = i\theta^{\mu\nu}\partial_\nu$. Therefore, as we can easily see the nonplanar part vanishes in $m_0\to\infty$ limit.

Now let us turn to the mixed contribution in which involves the $S_1(k;s,s')$ and $S_2(l;s,s')$ propagators, as well as the $S_2(k;s,s')$ and $S_1(l;s,s')$ ones. However, we can easily verify that these contributions are the same when we use the cyclic property of the trace. Thus, after we apply the two-dimensional Dirac delta functions, the expression (\ref{Sef4}) becomes
\be
S_{12} = \frac{e^2}2\int\frac{d^3p}{(2\pi)^3}\frac{dp'_2}{2\pi}\,A_\mu(p_a,p_2)A_\nu(-p_a,p'_2)\,\Pi^{\mu\nu}_{12},
\en
where
\ba\label{Pi12}
\Pi^{\mu\nu}_{12} &=& m_0\,\tr\int\frac{d^3k}{(2\pi)^3}\int ds\,e^{-is(k_2-p_2)}\sech(m_0s) \int ds'e^{-is'(-k_2-p'_2)}\sech(m_0s') \\
&\times& \gamma^\mu \frac{\Slash{k}+M(k;s,s')}{k^2-m_0^2}\gamma^\nu(1+i\gamma^2)\frac{(k_a-p_a)\gamma^a}{(k_0-p_0)^2-(k_1-p_1)^2} \sin(p_a\!\wedge\! k_a+2p_a\!\wedge\! k_2) \nonumber\\
&\times&\sin[p_a\!\wedge\! k_a-p'_2\!\wedge\!(2k_a+p_a)] \nonumber.
\ea
Here the index $a$ denotes the fact that only the $0,1$ components of the corresponding vector are to be taken into account. To estimate the behavior of this expression in the $m_0\to\infty$ limit we approximate $M(k;s,s')\sim m_0\1$ and $\sech(m_0s)\simeq 2e^{-|m_0||s|}$. In this approximation, we can integrate over $s$ and $s'$,
\be
\int ds\,e^{-is(k_2-p_2)}\sech(m_0s) = \frac{4m_0}{(k_2-p_2)^2+m_0^2},
\en
and the Eq. (\ref{Pi12}) can be rewritten as
\ba
\Pi^{\mu\nu}_{12} &=& m_0\,\tr\int\frac{d^3k}{(2\pi)^3} \gamma^\mu \frac{\Slash{k}+m_0}{k^2-m_0^2}\gamma^\nu(1+i\gamma^2)\frac{(k_a-p_a)\gamma^a}{(k_0-p_0)^2-(k_1-p_1)^2}\frac{4m_0}{(k_2-p_2)^2+m_0^2}\nonumber\\ &\times&\frac{4m_0}{(k_2+p'_2)^2+m_0^2}  \sin(p_a\!\wedge\! k_a+2p_a\!\wedge\! k_2)\sin[p_a\!\wedge\! k_a-p'_2\!\wedge\!(2k_a+p_a)].
\ea
Now, in order to integrate over $k_2$ in the above expression we first use Feynman parameters to combine the denominators and thus we get
\ba\label{Pi12_2}
\Pi^{\mu\nu}_{12} &=& 32m_0^3\int_0^1dx\int_0^{1-x}dy\;\tr\int\frac{d^3k}{(2\pi)^3} \frac{\gamma^\mu[k_a\gamma^a+(k_2+xp_2-yp'_2)\gamma^2+m_0]}{(k_2^2+\Delta_0^2)^3[(k_0-p_0)^2-(k_1-p_1)^2]} \\
&\times& \gamma^\nu(1+i\gamma^2)(k_a-p_a)\gamma^a\sin(p_a\!\wedge\! k_a+2p_a\!\wedge\! k_2)\sin[p_a\!\wedge\! k_a-p'_2\!\wedge\!(2k_a+p_a)] \nonumber
\ea
with $\Delta_0^2 = m_0^2+x(1-x)p_2^2+y(1-y)p^{\prime 2}_2+2xyp_2p'_2-(1-x-y)(k_0^2-k_1^2)$. Finally, we can use the nonplanar integral
\be
\int\frac{dk_2}{2\pi}\frac{e^{ik_2\tilde p_2}}{(k_2^2+\Delta_0^2)^3} = \frac{1}{16\Delta_0^3}\left(\tilde p_2^2 + \frac{3|\tilde p_2|}{\Delta_0}+\frac{3}{\Delta_0^2}\right)e^{-\Delta_0|\tilde p_2|},
\en
where $\tilde p_2=\theta_{2b}p^b$. Note that this expression displays an exponential decay as $\Delta_0$ grows, and therefore in the limit $m_0 \to \infty$ we can conclude that this contribution vanishes. In the fundamental representation this contribution also vanishes because the phase factor which involves the $k_2$ integration is the same.

To conclude our analysis we will consider the term involving the $S_2(k;s,s')$ and $S_2(l;s,s')$ propagators. The contribution to the effective action (\ref{Sef4}) takes the form
\ba\label{Sef22}
S_{22} &=& \frac{m_0^2e^2}{8} \int \frac{d^3p}{(2\pi)^3}\frac{dp'_2}{2\pi} \int ds\, ds'\, \sech^2(m_0s)\sech^2(m_0s')  \nonumber\\
&\times& \Pi_{22}^{\mu\nu} A_\mu(p_a,p_2)\,e^{isp_2} A_\nu(-p_a,p'_2)\,e^{is'p'_2}
\ea
with the polarization tensor given by
\ba\label{Pi22}
\Pi_{22}^{\mu\nu} &=& \tr\int\frac{d^2k}{(2\pi)^2} \gamma^\mu(1+i\gamma^2)\frac{k_a\gamma^a}{k^2}\gamma^\nu(1+i\gamma^2)\frac{(k_b-p_b)\gamma^b}{(k-p)^2} \sin(p_a\!\wedge\! k_a) \nonumber\\
&\times&\sin[p_a\!\wedge\! k_a+i\partial_{s'}\!\wedge\! (2k_a-p_a)],
\ea
where now the square of the vector is taken in the two-dimensional space. Using the inverse Fourier transformation with respect to the arguments $p_2$ and $p'_2$, the expression (\ref{Sef22}) can be rewritten in the form of an integral in which all momenta are clearly two-dimensional, yielding
\be\label{Sef22_1}
S_{22} = \frac{m_0^2e^2}{8} \int \frac{d^2p}{(2\pi)^2}\int ds\, ds'\, \sech^2(m_0s)\sech^2(m_0s')\, \Pi_{22}^{\mu\nu} A_\mu(p_a,s) A_\nu(-p_a,s').
\en
As in the limit $m_0 \to \infty$ we have
\be
m_0\,\sech^2(m_0s) = 2\delta(s),
\en
we can integrate over $s$ and $s'$, so that the above expression can be rewritten as
\be
S_{22} = \frac{e^2}2 \int \frac{d^2p}{(2\pi)^2} A_\mu(p_a) A_\nu(-p_a) \Pi_{22}^{\mu\nu}.
\en
The expression (\ref{Pi22}) can be represented in the form
\be\label{Pi22_1}
\Pi^{\mu\nu}_{22} = \Gamma^{\mu a \nu b}\,\int\frac{d^2k}{(2\pi)^2} \frac{k_a}{k^2}\frac{(k_b-p_b)}{(k-p)^2} \sin^2(p_a\!\wedge\! k_a),
\en
where
\be
\Gamma^{\mu a\nu b} = \tr\,\gamma^\mu(1+i\gamma^2)\gamma^a\gamma^\nu(1+i\gamma^2)\gamma^b
\en
(in the fundamental representation the phase factors also mutually cancel for this contribution). A straightforward calculation of the trace shows that $\Gamma^{\mu a\nu b}$ vanishes if at least one of the indices $\mu,\nu$ is equal to 2. Therefore we restrict ourselves to the object $\Gamma^{cadb}$ with all indices taking values $0,1$, so that the calculation of $\Gamma^{cadb}$ yields  
\be
\Gamma^{cadb} = 4(g^{ad}-\epsilon^{ad})(g^{bc}-\epsilon^{bc}),
\en
with $\epsilon^{ab}$ being the two-dimensional Levi-Civita symbol, with $\epsilon^{01}=1$. The important property of $\Gamma^{cadb}$ is that $g_{ab}\Gamma^{cadb}=0$. Now, in order to evaluate the integrals over $k_0$ and $k_1$ in the expression (\ref{Pi22_1}) we can use the Feynman parameter, so that
\ba
\int\frac{d^2k}{(2\pi)^2} \frac{k_a(k_b-p_b)}{k^2(k-p)^2} \sin^2(p_a\!\wedge\! k_a) &=& \frac12 \int_0^1dx \int\frac{d^2k}{(2\pi)^2} \frac{k_ak_b-x(x-1)p_ap_b}{[k^2-x(x-1)p^2]^2} \nonumber\\ &\times&\left[1-\cos(2p_a\!\wedge\! k_a)\right].
\ea
Thus, using the conventional Feynman integrals and the nonplanar ones, given by
\ba
\int\frac{d^2k}{(2\pi)^2} \frac{e^{ik_a\tilde p^a}}{(k^2-\Delta^2)^2} &=& \frac{i}{4\pi} \frac{|\tilde p|}{\Delta} K_1(\Delta|\tilde p|), \\
\int\frac{d^2k}{(2\pi)^2} \frac{k_ak_be^{ik_a\tilde p^a}}{(k^2-\Delta^2)^2} &=& -\frac{i}{4\pi}\,K_0\left(\Delta|\tilde p|\right)g_{ab} + \frac{i}{4\pi}\Delta|\tilde p| \,K_1\left(\Delta|\tilde p|\right) \frac{\tilde p_a \tilde p_b}{\tilde p^2},
\ea
where $\tilde p_a=\theta_{ab}p^b$, $\Delta^2 = x(x-1)p^2$, and $K_0$ and $K_1$ are the modified Bessel functions, we have
\be
\Pi^{cd}_{22} = -\frac{i}{8\pi} \int\frac{d^2p}{(2\pi)^2}\left[\frac{p_ap_b}{p^2} - \int_0^1dx\Delta|\tilde p| \,K_1\left(\Delta|\tilde p|\right) \left(\frac{p_ap_b}{p^2}+\frac{\tilde p_a\tilde p_b}{\tilde p^2}\right) \right] \Gamma^{cadb}.
\en
Defining $\theta_{ab}=\theta\epsilon_{ab}$, we can simplify the $\theta$ dependence of the coefficient 
\be\label{coef}
\left(\frac{p_ap_b}{p^2}+\frac{\tilde p_a\tilde p_b}{\tilde p^2}\right) = \left(\frac{p_ap_b}{p^2}+\frac{\epsilon_{al}\epsilon_{bm}p^lp^m}{p^2}\right),
\en 
which arose in the nonplanar contribution. However, as $\epsilon_{al}\epsilon_{bm}p^lp^m=g_{ab}p^2-p_ap_b$ this nonplanar contribution vanishes for any choice in $\theta$. Thus, returning to the space of coordinates, we can write down the chiral expression for the effective action (\ref{Sef22_1}) in the following form
\be
S_\mathrm{chir} = \frac{ie^2}{2\pi}\int d^2z A_c\,\epsilon^{ad}\,\frac{\partial_a\partial^c}{\partial^2}\,A_d.
\en

Let us now show that the gauge variation of the above expression cancels the gauge variation of the planar contribution of Chern-Simons action (\ref{CS}) exactly. We then have
\be
\delta S_\mathrm{chir} = \frac{ie^2}{2\pi}\int d^2z \left(\delta A_c\,\epsilon^{ad}\,\frac{\partial_a\partial^c}{\partial^2}\,A_d+ A_c\,\epsilon^{ad}\,\frac{\partial_a\partial^c}{\partial^2}\,\delta A_d\right),
\en
where $\delta A_a = \partial_a\Lambda$. Substituting this and also using the antisymmetry of $\epsilon^{ab}$, we get
\be\label{deltachir}
\delta S_\mathrm{chir} = -\frac{ie^2}{2\pi}\int d^2z\,\Lambda\,\epsilon^{ab}\,\partial_a A_b.
\en
On the other hand, after we take into account the $m_0 \to \infty$ limit in (\ref{CS}) the gauge variation becomes
\be\label{deltaCS}
\delta S_\mathrm{CS} = -\frac{ie^2}{4\pi}\int d^3z\, \mathrm{sgn}(s) \epsilon^{\mu\nu\lambda} \partial_\mu\Lambda \partial_\nu A_\lambda = \frac{ie^2}{2\pi}\int d^3z\,\delta(s)\,\Lambda\,\epsilon^{ab}\,\partial_a A_b,
\en
where we have substituted $\mathrm{sgn}(m_0)$ for $\mathrm{sgn}(s)$. Hence, the sum of the planar contributions (\ref{deltachir}) and (\ref{deltaCS}) is zero, whereas the nonplanar contributions vanish either in the $m_0 \to \infty$ limit in (\ref{CS}) or precisely through (\ref{coef}).  

\section{Summary}

We have studied the possibility of the appearance of the  chiral anomaly in the noncommutative spinor electrodynamics with a fermion domain wall mass. It turns out that, in the case where the spinors are coupled to gauge field via the fundamental representation, the induced terms are purely planar and  coincide with the ones for the commutative counterpart of this theory, i.e., the anomaly is exactly cancelled. In the case of the adjoint representation the nonplanar part of two-dimensional contribution gives a nontrivial integrand   which vanishes  upon integration; on the other hand, the three dimensional nonplanar part only vanishes in the limit $m_0\to \infty$ . Then, in either case, in the limit of small $\theta$ the Callan-Harvey effect is not affected by the noncommutativity of the space-time.

{ In the context of the noncommutative field theory, this has the following natural interpretation. It is well known that the noncommutative field theories, in general, are characterized by the UV/IR mixing implying in a highly nontrivial low-energy effective dynamics. In particular, the arising of quadratic or linear UV/IR infrared divergences would generate  new kinds of contributions to the effective action, which, being proportional to $1/\theta^2$ or to $1/\theta$ respectively, in principle could generate anomalies. Thus, the absence of anomalies in this model in the small $\theta$ limit is a natural consequence of the absence of  dangerous UV/IR infrared divergences.}

\vspace{1cm}

{\bf Acknowledgements.} Authors are grateful to Prof. D. Bazeia for useful discussions. This work was partially supported by Funda\c{c}\~{a}o de Amparo \`{a} Pesquisa do Estado de S\~{a}o Paulo (FAPESP) and Conselho Nacional de Desenvolvimento Cient\'{\i}fico e Tecnol\'{o}gico (CNPq). The work by T. M. has been supported by FAPESP, project 06/06531-4. The work by A. Yu. P. has been supported by CNPq-FAPESQ DCR program, CNPq project No. 350400/2005-9.


\begin{thebibliography}{99}
\bibitem{CH} C. G. Callan, Jr. and J. A. Harvey, Nucl. Phys. {\bf B250}, 427 (1985).
\bibitem{Naculich} S. Naculich, Nucl. Phys. {\bf B296}, 837 (1988).
\bibitem{Chand} S. Chandrasekharan, Phys. Rev. {\bf D49}, 1980 (1994).

\bibitem{Fos} G. D. Fosco, A. Lopez, Nucl. Phys. {\bf B538}, 685 (1999), hep-th/9807217; G. D. Fosco, A. Lopez, F. Schaposnik, Nucl. Phys. {\bf B582}, 716 (2000), hep-th/9912285.
\bibitem{dw} E. Fradkin, G. D. Fosco, A. Lopez, Phys. Lett. {\bf B451}, 31 (1999), hep-th/9902065; A. Rebhan, P. van Nieuwenhuizen, R. Wimmer, New J. Phys. {\bf 4}, 31 (2002), hep-th/0203137; Nucl. Phys. {\bf B648}, 174 (2003), hep-th/0207051.
\bibitem{FosTor} C. D. Fosco, G. Torroba, Phys. Lett. {\bf B620}, 174 (2005), hep-th/0505002.
\bibitem{fen} T. L. Ho, J. Fulco, J. R. Schrieffer, F. Wilczek, Phys. Rev. Lett. {\bf 52}, 1524 (1984); D. Boyanovsky, E. Dagotto, E. Fradkin, Nucl. Phys. {\bf B285}, 340 (1987); M. Stone, A. Garg, P. Muzikar, Phys. Rev. Lett. {\bf 55}, 2328 (1985); A. W. W. Ludwig, M. P. A. Fisher, R. Shankar, G. Grinstein, Phys. Rev. {\bf B50}, 7526 (1994).
\bibitem{NCFT} R. J. Szabo, Phys. Rep. {\bf 378}, 207 (2003) [hep-th/0109162]; M. R. Douglas and N. A. Nekrasov, Rev. Mod. Phys. {\bf 73}, 977 (2001) [hep-th/0106048].
\bibitem{Minw} S. Minwalla, M. van Raamsdonk, N. Seiberg, JHEP {\bf 02}, 020 (2000) [hep-th/9912072].
\bibitem{Pres} P. Presnajder, J. Math. Phys. {\bf 41}, 2789 (2000) [hep-th/9912050].
\bibitem{Ard} F. Ardalan, N. Sadooghi, Int. J. Mod. Phys. {\bf A16}, 3151 (2001) [hep-th/0002143].
\bibitem{Mart} J. M. Gracia-Bondia, C. P. Martin, Phys. Lett. {\bf B479}, 321 (2000) [hep-th/0002171].
\bibitem{Bon} L. Bonora, M. Schnabl, A. Tomasiello, Phys. Lett. {\bf B485}, 311 (2000) [hep-th/0002210].
\bibitem{NCQED} M. Hayakawa, Phys. Lett. B {\bf 478}, 394 (2000) [hep-th/9912094].
\bibitem{Mariz}T.~Mariz, J.~R.~S.~Nascimento, R.~F.~Ribeiro and F.~A.~Brito, Phys.\ Rev.\  D {\bf 68}, 087701 (2003) [hep-th/0305003].
\end{thebibliography}
\end{document}